\documentclass[journal]{IEEEtran}

\ifCLASSINFOpdf

\else

\fi
\usepackage{mathtools,amssymb,lipsum}

\usepackage{amsmath}
\usepackage{array,hyperref}
\usepackage{graphicx}
\usepackage{amssymb}
\usepackage{xcolor}
\usepackage[normalem]{ulem}
\usepackage[ruled,vlined]{algorithm2e}
\usepackage{amsfonts}
\usepackage{amsthm}

\DeclareMathOperator{\EX}{\mathbb{E}}
\DeclareMathOperator*{\argmax}{argmax}

\begin{document}

\title{Generalized State-Feedback Controller Synthesis for Underactuated Systems through Bayesian Optimization}

\author{
Miguel~A.~Solis,~\IEEEmembership{Member,~IEEE,} and~Sinnu~Susan~Thomas,~\IEEEmembership{Member,~IEEE}
\thanks{M. A. Solis is with Automation and Robotics Department, Faculty of Engineering, Universidad Andres Bello, Santiago, Chile. e-mail: miguel.solis@unab.cl}
\thanks{Sinnu S. Thomas is with Machine Learning and Optimization Lab, School of Computer Science and Engineering, Digital University Kerala (IIITMK) Kerala India. e-mail: sinnu.thomas@iiitmk.ac.in}
\thanks{Manuscript received March 31, 2021.}}

%
%

\markboth{UNDER REVIEW}%
{Solis \MakeLowercase{\textit{et al.}}: Generalized State-Feedback Controller Synthesis for Underactuated Systems through Bayesian Optimization}

\maketitle

\begin{abstract}
Underactuated systems pose the challenge of being able to control a plant whose degrees of freedom are not necessarily directly linked to an actuator or where such a relationship is not straightforward. Rotary inverted pendulum is an example of such systems, which on its simplest representation consists of a pendulum whose vertical angle should be taken up to the upward unstable position based on the impulse given from another bar and an appropriate control strategy, bar that is controlled by an electrical motor. This problem is often tackled by linear control theory with state-feedback controllers that is frequently obtained by means of designing a feedback gain meeting some constraints. This article reports a Bayesian Optimization approach for designing a generalized state-feedback controller, that involves more parameters than a simple state-feedback control law, but with the benefit of achieving lower control effort in terms of the signal amplitude. The source code is made publicly available to facilitate further research.
\end{abstract}


\IEEEpeerreviewmaketitle

\section{Introduction}

\IEEEPARstart{U}{nderactuated} plants pose a challenging problem for control design, since in addition to being able to reject disturbances and deal with model uncertainties, its corresponding controller has to generate a control signal such that a given state is taken to zero without necessarily requiring to be linked to an actuator.

Underactuated systems are appealing due to its interesting spectrum applications such as aerospace systems or robotic platforms \cite{olfati2000nonlinear,spong1998underactuated,birglen2007underactuated}, and the efforts of the controller may be diminished if it takes advantages from the structural dynamics often involved, instead of the classical feedback scheme where the controller aims to nullify the system to be controlled.

State feedback controllers are often appealing due to their structural simplicity since it consists only on designing a feedback gain. Nevertheless, when
stabilizing a given plant, its dynamics could lead to the gain of a static feedback controller to take higher values than desired translating into higher control efforts. On the other hand, a dynamic state feedback controller is capable of achieving the same or even better performance by introducing additional parameters into the model to be designed.

The rotary inverted pendulum is one of the most popular control experiments within the underactuated mechanical systems area, having a nonlinear model but allowing a linear approximation around the upper unstable position, broadening the extent of possible controller classes to the well-known and extensively studied linear realm. This plant consists of a controlled arm on the horizontal plane rotating around a central axis, and a pendulum linked to one of the ends of this arm, rotating over the vertical plane. This plant allows the study of nonlinear dynamics of simplified models that are also possible to construct physically, building mechatronic platforms with enhanced reproducibility features.

A well-known linear control problem is the Linear Quadratic Regulator (LQR), that consists on the assumption of the system to be linear, and the performance index is given in terms of a quadratic function \cite{anderson2007optimal}. LQR is particularly appealing given that its solution is obtained by solving an algebraic Riccati equation (ARE). In the other hand, the linear quadratic tracking (LQT) problem also assumes a linear model for the process dynamics and a quadratic function for the performance index, but the main objective is to design a controller such that the measured output of the process to be controlled, follows an exogenous reference signal, so the LQR could be considered as a particular case of the LQT problem.

State-feedback controllers are attractive due to its simple model, that consists on a gain whose input is the state of the system and as a consequence, computation cost is minimized both for generating the corresponding control signal and features analysis such as closed-loop poles placement and speed of convergence for associated natural modes. Generalized state-feedback controller extends the structure of the linear model typically used for the plant, adding an additional degree of freedom on the controller at the cost of including more parameters to be designed.

The remainder of this article is as follows: Section \ref{sec:literature} shows related work in literature focused on state-feedback controllers and its applications on underactuated systems. Section \ref{sec:theoretical} presents the rotary inverted pendulum with its linear model obtained from a phenomenological analysis and its corresponding linearization, concluding with the linear quadratic regulator problem and classical state-feedback controller. Section \ref{sec:bayesian} shows the bayesian optimization algorithm applied for tuning the generalized state-feedback controller parameters, whose results are reported on Section \ref{sec:results}, followed by some final remarks and conclusion on Section \ref{sec:conclusion}.

 

\section{Literature Review}\label{sec:literature}

\subsection{Underactuated Systems}
Control community have done a great variety of work related to underactuated systems and introduced some benchmark problems like the cart-pole \cite{yu2008closed}, where the control problem requires not only to stabilize the pendulum in the upper unstable position, also displacement of the cart has to be considered, increasing design complexity. There are other similar underactuated mechanical systems with models of reduced order, allowing to capture the essence of a problem without introducing all the complexity frequently involved in real world applications. An example of a trivially underactuated system, with two degrees of freedom and one actuator, is the rotary inverted pendulum (RIP) also known as Furuta pendulum \cite{aastrom2000swinging}.

Different control techniques have been implemented on similar problems, such as \cite{hassanzadeh2011controller} that applies different evolutionary algorithms in order to adjust parameters of a PID controller, including genetic algorithms, particle swarm optimization and ant colony optimization that simulates behavior of ants and its ability to find the shortest path from their nest to a food source \cite{dorigo2006ant}. Although these methods may provide sufficient performance for the obtained controller, there are no guarantees of stability nor convergence speed on the intermediate process.

Other model-based control approaches on the Furuta pendulum have been also reported on literature, like \cite{azar2015adaptive} where the design and implementation of an adaptive sliding mode controller is discussed, along with other sliding mode variations in order to compare their performance within that class of controllers. Although nonlinear control does not deal with model approximations, linear control is a powerful approach that can provide a more in-depth stability analysis due to the extensively studied linear theory.

Controller design has also been tackled from adaptive approaches such as \cite{kong2019asymmetric} or \cite{arefi2014adaptive} that generates a control signal based on an artificial neural network based controller. Although those works are mainly based on a neural approach, artificial neural networks may also be used to enhance other learning approaches such as reinforcement learning algorithms \cite{sutton2018reinforcement} that have been extensively applied for solving the LQR problem, that is the regulator problem when the system is assumed to be linear, and the performance index is given in terms of a quadratic function \cite{anderson2007optimal}, is particularly appealing given that its solution is obtained by solving an algebraic Riccati equation (ARE). Then,
there algorithms that basically starts with an admissible control policy and then iterates between
given steps until variations on the policy or the specified value function are negligible, as seen on \cite{lewis2011reinforcement}. In the other hand, the linear quadratic tracking (LQT) problem also assumes a linear model for the process dynamics and a quadratic function for the performance index, but the main objective is to design a controller such that the measured output of the process to be controlled, follows an exogenous reference signal, so the LQR could be considered as a particular case of the LQT problem.

Nevertheless, the LQT has not received much attention in the literature mainly because for most reference signals the infinite horizon cost becomes unbounded \cite{barbieri2000infinite}. Work in \cite{qinzhang14} tackles the problem on the continuous time domain by solving an augmented ARE obtained from the original system dynamics and the reference trajectory dynamics, while \cite{kiumarsi2014reinforcement} take a similar approach for the discrete-time case, where a Q-learning algorithm \cite{watkins1992q} is obtained for solving the LQT problem without any model knowledge.

Although the above references are novel and sound when considering not requiring model knowledge, they often involve more computational effort. Generalized state-feedback controllers proposed in this document include three more parameters to be designed but with the benefit of achieving lower amplitude on the generated control signal, expecting to achieve stationarity with the same velocity than a classical state-feedback approach, confining the control signal to be smaller at the cost of introducing an additional natural mode on the closed-loop response.

On the linear class of controllers, work in \cite{ramirez2014linear} a linear controller based on a linear observer is obtained under an active disturbance rejection control scheme. A similar approach of the switched control strategy to be used on the RIP can be found on \cite{olivares2014switched}, where a energy-based nonlinear controller is designed for the swing-up task of the flywheel inverted pendulum, and also a locally stabilizing controller is obtained by means of tuning the appropriate parameters of a PID controller in order to take the pendulum from its stable rest position to the upward unstable vertical line.

When tackling the partially observable case, due to missing elements from the state vector, or given that there could be high costs associated with measuring each element from the resulting state, suitable estimations have to be made. Nevertheless and without detriment of contribution, this work assumes system state is directly achievable so it can be used on the control strategy given that there is full knowledge of the model of the plant, in particular, knowledge about variance of both process and measurement noise, and model parameters.

\subsection{Bayesian Optimization}
Bayesian Optimization (BO) creates a surrogate model of the black-box objective function and performs optimization by iterative sampling based on an acquisition function defined based on the surrogate model \cite{Mockus:1974:BMS:646296.687872,BrochuarXiv2010,NIPS2017_df1f1d20,ghoreishi2020bayesian,baptista2018bayesian,shahriari2015taking,179841,8315110}. BO has been studied in many fields for the optimization of unknown functions. The performance of the BO is highly dependent on the choice of acquisition function made to find the best-observed value. Nguyen et al.\ \cite{pmlr-v77-nguyen17a} proposed convergence criteria for these acquisition functions in order to avoid unwanted evaluations. It has been applied with great success in  machine learning applications \cite{Snoek2012NIPS}, analog circuits \cite{Lyu2018TCS}, voltage failures \cite{Hu2019DAC}, aerospace engineering \cite{Lam2018AIAA}, asset management \cite{Gonzalvez2019arXiv}, pharmaceutical products \cite{Sano2019JPI}, laboratory gas-liquid separator \cite{Kocijan2014IIM}, multi objective optimization \cite{Wang2019GECCO}, electron lasers \cite{pmlr-v97-kirschner19a}, and autonomous systems \cite{8303767}.

\section{Theoretical Framework} \label{sec:theoretical}

\subsection{Rotary Inverted Pendulum}
Consider a second order and controllable dynamical system given by:
\begin{equation}
\label{eq:din}
\mathbf{\ddot{q}} = f(\mathbf{q},\mathbf{\dot{q}},\mathbf{u},t),
\end{equation}
where $\mathbf{u}$ corresponds to the control vector, $\mathbf{q}$ and $\mathbf{\dot{q}}$ are the positions and velocities vector respectively, and $t$ denotes the possible influence of time over the acceleration vector $\mathbf{\ddot{q}}$. For the case that concerns this study, where dynamics are affine on the commanded torque, this expression can be rewritten as:
\begin{equation}
\mathbf{\ddot{q}} = f_1(\mathbf{q},\mathbf{\dot{q}},t) + f_2(\mathbf{q},\mathbf{\dot{q}},t)\mathbf{u}.
\end{equation}
Then, in formal terms a control system described by Eq.\eqref{eq:din} is called underactuated in the configuration $(\mathbf{q},\mathbf{\dot{q}},t)$ if it is not possible to drive an instantaneous acceleration on any arbitrary direction, i.e.:
\begin{equation}
\label{eq:subact}
rank(f_2(\mathbf{q},\mathbf{\dot{q}},t)) < dim(\mathbf{q}).
\end{equation}

\begin{figure}[!htbp]
\centering
\includegraphics[width=.5\textwidth]{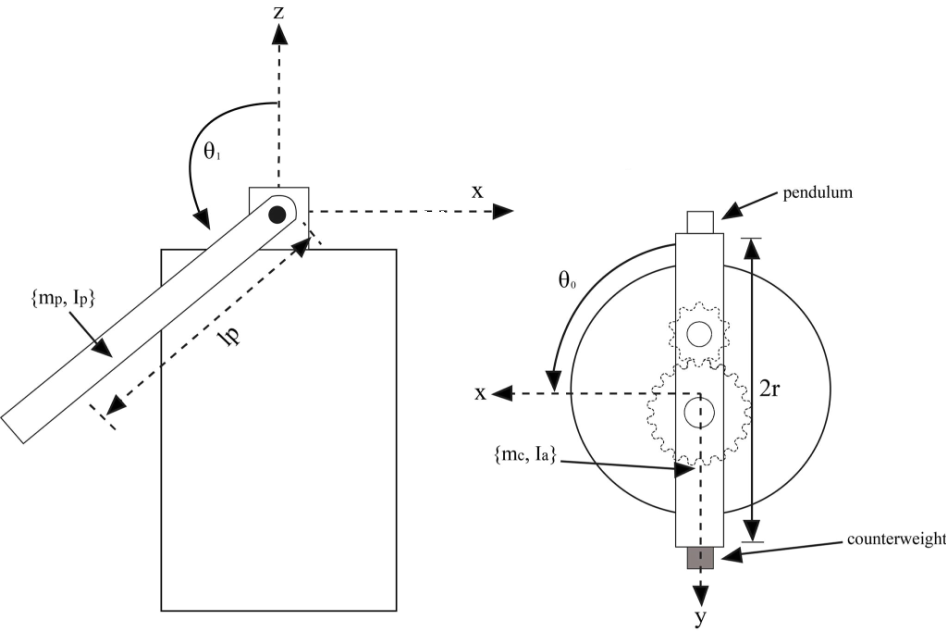}
\caption{Side and top view of Furuta pendulum}
\label{fig:furuta}
\end{figure}

The rotary inverted pendulum (RIP), also known as Furuta pendulum is an example of an underactuated system as illustrated in Fig. \ref{fig:furuta}, it consists of a controlled arm in the horizontal plane rotating around a central axis, and a pendulum linked to one of the ends of this arm, rotating over the vertical plane. This plant allows us to study nonlinear dynamics of simplified models that are also possible to construct physically, building mechatronic platforms with enhanced reproducibility features.

According to Fig. \ref{fig:furuta}, there is a DC motor responsible for controlling the arm position measured by $\theta_0$ through the armature voltage, and its main parameters are the armature resistance and inductance, $R_a$ and $L_a$ respectively.

\subsection{Model formulation}

We define $\{l_p,m_p,I_p,\theta_1\}$, where $l_p$ stands for the length from the rotating axis of the pendulum to its center of mass, $m_p$ denotes the mass of the pendulum with its moment of inertia $I_p$ and its rotation angle $\theta_1$. Then, $r$ corresponds to the arm radius that has a moment of inertia $I_a$, introducing a counterweight of mass $m_c$ that takes the center of the mass of the rotating arm at height $h$. Consider the rotation angle of the arm to be given by $\theta_0$. A detailed description of parameters involved in the model along with its values for the physical prototype is shown in Table \ref{tab:parameters}, that comes from a phenomenological analysis and matches the physical prototype documented on \cite{solis2019switched}.

The Lagrangian, $L(\mathbf{\theta},\mathbf{\dot{\theta}})$, is then given by Eq. \eqref{eq:lagrangian}, where $E_k$ and $E_p$ denote the kinetic and potential energy respectively, while $\mathbf{\theta}$ is the generalized coordinates vector, $\mathbf{\theta} = \left[\begin{array}{cc}\theta_0 & \theta_1\end{array}\right]^{\intercal}$.

\begin{equation}\label{eq:lagrangian}
L(\mathbf{q},\mathbf{\dot{q}}) = E_k(\mathbf{q},\mathbf{\dot{q}}) - E_p(\mathbf{q},\mathbf{\dot{q}})
\end{equation}

Kinetic energy for the pendulum is given by the sum of translational and rotational components, while kinetic energy for the arm is given by rotation and tangential kinetic energy:
\begin{equation}
E_k = \frac{1}{2}\hat{J}_0\dot{\theta}_0^2 + \frac{1}{2}\hat{J}_1\dot{\theta}_1^2 + \frac{1}{2}m_pl_p^2\dot{\theta}_0^2sin^2(\theta_1) - m_prl_p\dot{\theta}_0\dot{\theta}_1cos(\theta_1),
\end{equation}
where $\theta_0$, $\theta_1$ are the angular velocities for the arm and pendulum respectively.

And $\hat{J}_0$ and $\hat{J}_1$ are given by
\begin{subequations}
\begin{align}
\hat{J}_0 &= I_a + r^2 (m_p + m_c),\\
\hat{J}_1 &= I_p + m_p l_p^2,
\end{align}
\end{subequations}
while potential energy is given in terms of the pendulum and counterweight masses:
\begin{equation}
E_p = m_pgl_pcos(\theta_1) + m_cgh.
\end{equation}

\begin{table}[!htbp]
\centering
\begin{tabular}{|c|c|c|}
\hline 
\textbf{Symbol} & \textbf{Description} & \textbf{Value} \\ 
\hline 
$m_p$ & Pendulum mass & $0.1$ $kg$ \\ 
\hline 
$m_c$ & Counterweight mass & $0.01$ $kg$ \\ 
\hline 
$I_p$ & Pendulum inertial moment & $5.1\times 10^{-4}$ $kg$ $m^2$ \\ 
\hline 
$I_a$ & Arm inertial moment & $3.1\times 10^{-3}$ $kg$ $m^2$ \\ 
\hline 
$r$ & Arm radius & $0.13$ $m$ \\ 
\hline 
$l_p$ & Pendulum mass center & $0.125$ $m$ \\ 
\hline 
$h$ & Arm center of mass height & $0.055$ $m$ \\ 
\hline 
$C_0$ & Arm friction coefficient & $10^{-4}$\\
\hline
$C_1$ & Pendulum friction coefficient & $10^{-4}$\\
\hline
$R_a$ & Armature resistor & $8$ $\Omega$ \\
\hline 
$L_a$ & Motor inductance  & $10$ $mH$ \\
\hline 
$I_m$ & Motor inertia & $1.9\times 10^{-6}$ $kg$ $m^2$ \\
\hline 
$M_f$ & Motor mutual inductance & $0.0214$ $N$ $m/A$ \\
\hline 
$K_g$ & Gear reduction coefficient & $59927$ \\
\hline 
$K_{eg}$ & External gear reduction coefficient & $16$ \\
\hline 
$g$ & Gravitational acceleration & $9.806$ $m/s^2$ \\ 
\hline 
\end{tabular} 
\caption{RIP parameters}
\label{tab:parameters}
\end{table}

Then, according to the Euler-Lagrange equation:
\begin{equation}
\frac{\partial}{\partial t}\left(\frac{\partial L}{\partial\dot{q}_i}\right)-\frac{\partial L}{\partial q_i} = \tau_i, \quad i = \theta_0,\theta_1
\end{equation}
with $\tau_i$ being the moments applied to each coordinate, leading to

\begin{equation}
M_1(\theta_1)\ddot{q} + M_2(\theta_1)\dot{q}+M_3(\theta_1) = T,
\end{equation}
where $M_1$, $M_2$, and $M_3$ corresponds to inertia matrix, centripetal, and Coriolis torque matrices respectively and $T$ is the torque vector:
\begin{subequations}\label{eq:matrices}
\begin{align}
M_1(\theta_1) &= \left[\begin{array}{cc}\hat{J}_0+m_pl_p^2sin^2(\theta_1) & -m_prl_pcos(\theta_1)\\ -m_prl_pcos(\theta_1) & \hat{J}_1\end{array}\right],\\
M_2(\theta_1) &= \left[\begin{array}{cc}m_pl_p^2\dot{\theta}_1sin(2\theta_1)+C_0 & m_prl_p\dot{\theta}_1sin(\theta_1)\\ -\frac{1}{2}m_pl_p^2\dot{\theta}_0sin(2\theta_1) & C_1\end{array}\right],\\
M_3(\theta_1) &= \left[\begin{array}{cc}0\\-m_pgl_psin(\theta_1) \end{array}\right],\\
T &= \left[\begin{array}{cc}\tau_l\\0\end{array}\right],
\end{align}
\end{subequations}

with $C_0$ and $C_1$ as the friction coefficients of the arm and pendulum respectively, with values given in Table \ref{tab:parameters}. From Eqs. \eqref{eq:matrices}, the reader can note that the motor actuates just on the arm and matrices model the movement transfer to the pendulum. This load torque, $\tau_l$ is related to the motor dynamics that is related with electrical torque, $\tau_e$, and electrical angular velocity, $\omega_e$, given by
\begin{subequations}
\begin{align}\label{eq:vmotor}
V(t) &= R_a i(t) + L_a\frac{d}{dt}i(t) + M_f\omega_e,\\ \label{eq:tau_e}
\tau_e &= M_fK_g i(t),\\ \label{eq:taudiff}
\tau_e - \tau_l &= I_m\frac{d}{dt}\omega_e,
\end{align}
\end{subequations}
with parameters as described in Table \ref{tab:parameters}, and $V(t)$ and $i(t)$ are the applied voltage and current to the motor at time $t$, respectively.

Finally, applied torque and angular velocity is related with its corresponding electrical variables as follows:
\begin{subequations}
\begin{align}\label{eq:we}
\omega &= K_g^{-1}\omega_e,\\
\tau &= K_g\tau_e.
\end{align}
\end{subequations}

Then, the RIP model including motor dynamics is given by:
\begin{equation}\label{eq:ripwmotor}
\ddot{q} = \bar{M_1}(\theta_1)^{-1}\left(\bar{T}-\bar{M_2}(\theta_1)\dot{q}-M_3(\theta_1)\right),
\end{equation}
where new matrices are described by:
\begin{subequations}
\begin{align}
\bar{M_1}(\theta_1) &= \left[\begin{array}{cc}\hat{J}_0+m_pl_p^2sin^2(\theta_1)+K_g^2I_m & -m_prl_pcos(\theta_1)\\ -m_prl_pcos(\theta_1) & \hat{J}_1\end{array}\right],\\
\bar{M_2}(\theta_1) &= \left[\begin{array}{cc}m_pl_p\dot{\theta}_1sin(2\theta_1)+C_0 & m_prl_p\dot{\theta}_1sin(\theta_1)\\ -\frac{1}{2}m_pl_p\dot{\theta}_0sin(2\theta_1) & C_1\end{array}\right],\\
\bar{T} &= \left[\begin{array}{cc}M_fK_gi\\ 0\end{array}\right].
\end{align}
\end{subequations}

Define the state vector $\mathbf{x} = \left[\begin{array}{ccccc}\theta_0 &\dot{\theta}_0 & \theta_1 & \dot{\theta}_1 & i\end{array}\right]^{\intercal}$. Then, in order to obtain a linearized model around the equilibrium point $x = \left[\begin{array}{ccccc}0&0&0&0&0\end{array}\right]$, recall Eq. \eqref{eq:vmotor} and solve for the derivative of current $i$:
\begin{equation}\label{eq:didt}
\frac{d}{dt}i(t) = \frac{1}{L_a}V(t) - \frac{R_a}{L_a}i(t) - \frac{M_f}{L_a}\omega_e.
\end{equation}

From Eq. \eqref{eq:we} and noting that this prototype has an external gear reduction system that relates the arm angular velocity with its corresponding variable obtained from the motor described by
\begin{equation}\label{eq:thetaomega}
\dot{\theta}_0 = K_{eg}^{-1}\omega,
\end{equation}
Eq. \eqref{eq:didt} can be rewritten as
\begin{equation}
\label{eq:didt2}
\frac{d}{dt}i(t) = \frac{1}{L_a}V(t) - \frac{R_a}{L_a}i(t) - \frac{M_f}{L_aK_g}\dot{\theta}_0.
\end{equation}
Then, the following linear model is obtained
\begin{equation}\label{eq:linear}
\dot{x}(t) = Ax(t) + Bu(t) + v(t),
\end{equation}
\begin{equation}\label{eq:outputmodel}
    y(t) = Cx(t) + Du(t) + w(t),
\end{equation}
where $x$ is the previously defined state vector, $u$ corresponds to the control signal (applied voltage), while $v$ and $w$ are uncorrelated (Gaussian) zero-mean process and measurement noise respectively, with constant variances $P_v$ and $P_w$ accordingly. Matrices $A$ and $B$ are given by
\begin{table*}
\centering
\begin{minipage}{0.75\textwidth}
\begin{subequations}\label{eq:AandB}
\begin{align}
A &= \alpha\cdot\left[\begin{array}{ccccc}0 & \frac{1}{\alpha} &0 &0 &0\\ \\
0 & -\hat{J}_1C_0 & m_p^2l_p^2gr &m_prl_pC_1&M_fK_g\hat{J}_0\hat{J}_1\\ \\
0 & 0& 0 & \frac{1}{\alpha} &0\\ \\
0 & m_prl_pC_0 & \hat{J}_0m_pgl_p&-\hat{J}_0C_1&K_gM_fm_prl_p\\ \\
0 & -\frac{M_f}{L_aK_g\alpha} &0 &0&-\frac{R_a}{L_a\alpha}\end{array}\right],\\
B &= \left[\begin{array}{ccccc}0&0&0&0&\frac{1}{L_a}\end{array}\right]^{\intercal},
\end{align}
\end{subequations}
\medskip
\end{minipage}
\end{table*}
with $\alpha$ given by the following expression:
\begin{equation}
\alpha = \frac{1}{\hat{J}_0\hat{J}_1-(m_prl_p)^2}.
\end{equation}
Finally, noting that pendulum angles $\theta_0$ and $\theta_1$ and motor current $i$ are the measurable outputs, matrices $C$ and $D$ on \eqref{eq:outputmodel} becomes:

\begin{subequations}\label{eq:CandD}
\begin{align}
C &= \left[\begin{array}{ccccc}1 & 0 &0 &0 &0\\ \\
0 & 0 & 1 & 0 &0\\ \\
0 & 0& 0 & 0 &1\end{array}\right],\\
D &= 0.
\end{align}
\end{subequations}
 Matrices $(A,B,C,D)$ are the state space variables \cite{friedland2012control} with dimensions $A \in \mathbb{R}^{n_x\times n_x}$, $B \in \mathbb{R}^{n_x\times n_u}$ and $C \in \mathbb{R}^{n_y\times n_x}$, where $n_x$, $n_u$, and $n_y$ stands for the number of states, number of control signals, and number of outputs respectively.
 
\subsection{Linear Quadratic Regulator}

Considering a system described in state space variables with linear model in Eq. \eqref{eq:linear} and \eqref{eq:outputmodel}, the linear quadratic regulator \cite{anderson2007optimal} consists on a state-feedback controller such as described on Fig. \ref{fig:sf_scheme_classical}, that defines a cost function to be minimized and given by

\begin{equation}\label{eq:costlqr}
    J = \int_0^{\infty}{\left(x^{\intercal}(t)Qx(t) + u^{\intercal}(t)Ru(t)\right)dt},
\end{equation}

where $Q$ and $R$ are weighting positive definite and positive semi-definite matrices respectively.  Then, state-feedback control law that minimizes Eq. \eqref{eq:costlqr} is given by

\begin{equation}\label{eq:lqr}
    u(t) = -Kx(t),
\end{equation}

where $K$ is such that

\begin{equation}
    K = R^{-1}B^{\intercal}P,
\end{equation}

and $P$ stands for the solution of the algebraic Ricatti equation (ARE):

\begin{equation}\label{eq:are}
    A^{\intercal}P + PA - PBR^{-1}B^{\intercal}P + Q = 0.
\end{equation}
 
\section{Controller Synthesis} \label{sec:controller}

\begin{figure}
\centering
\includegraphics[width=0.5\textwidth]{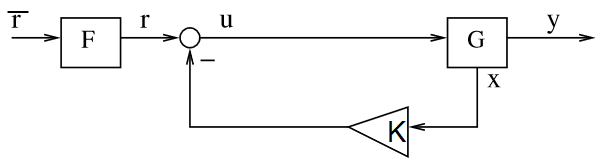}
\caption{Classical state-feedback control scheme}
\label{fig:sf_scheme_classical}
\end{figure}

The plant $G$, at the moment is assumed to be of the form as described in Eq. \eqref{eq:AandB} and \eqref{eq:CandD}. Consider the classical state-feedback control scheme shown in Fig. \ref{fig:sf_scheme_classical}, that corresponds to the linear quadratic regulator on Eq. \eqref{eq:lqr}, where $K$ corresponds to a (matrix) gain that weights the plant state $x(t)$ in order to generate a control signal $u(t)$ for tracking reference $r(t)$ at time $t$. Usually, on a state-feedback control scheme as shown in Fig. \ref{fig:sf_scheme_classical}, $r(t)$ is a pre-filtered reference that translates a reference $\bar{r}(t) \in \mathbb{R}^{n_y}$ that is the desired value that should take $y(t)$ at time $t$, into a suitable vector with appropriate dimensions.

A stabilizing controller (matrix $K \in \mathbb{R}^{n_u\times n_x}$) for the plant $G$, has to be designed such that the eigenvalues of $eig(A-BK)$ lie on the left half-plane on the continuous-time case, which on the discrete-time case translates to the region inside the unit circle \cite{anderson2007optimal}. Moreover, if we want to focus also on tracking, that is getting $y(t)$ as close as possible to $\bar{r}(t)$, the prefilter $F$ would be given by
\begin{equation}
F = -\left(C\left(A-BK-I\right)^{-1}B\right)^{-1}.
\end{equation}

\begin{figure}[!htbp]
\centering
\includegraphics[width=0.5\textwidth]{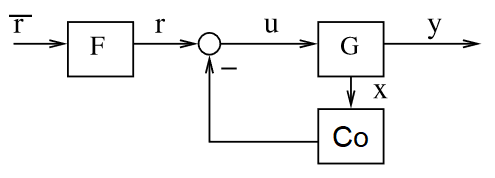}
\caption{State-feedback control scheme}
\label{fig:sf_scheme}
\end{figure}

We consider a state-feedback control scheme to design a controller such that it has its own dynamics as shown in Fig. \ref{fig:sf_scheme},

where we will consider that the reference vector $r(t)$ is a pre-filtered version of $\bar{r}(t)$, resulting on a $n_u$-elements vector representing the desired output $y(t)$ that is contained on $\bar{r}(t)$. Then, the controller is synthesized on $Co$ from Fig. \ref{fig:sf_scheme}, and assuming linearity, it can be described on a state model by
\begin{subequations}\label{eq:controller}
\begin{align}
\dot{x}_c(t) &= A_cx_c(t) + B_cx(t),\\
u(t) &= r(t) - \left(C_cx_c(t) + D_cx(t)\right),
\end{align}
\end{subequations}
where $c$ subscript is set to stress the difference between matrices $(A,B,C,D)$ from the plant model to the $(A_c,B_c,C_c,D_c)$ from the controller, as well as the state of the plant, $x(t)$ to the internal state of the controller itself, $x_c(t)$.

Once the controller has been designed, the pre-filter $F$ for transforming reference $\bar{r}(t)$ into $r(t)$ should be chosen such that the transfer function from $\bar{r}(t)$ to $y(t)$ is unitary, in order to ensure stationary tracking. Indeed, prefilter $F$ is given by
\begin{equation}
F = -\left(C\left(A-I-BD_c+BC_c\left(A_c-I\right)^{-1}B_c\right)^{-1}B\right)^{-1}
\end{equation}

It can be seen from Eq. \eqref{eq:controller}, that when $C_c = 0$ we have exactly the same state-feedback law as in the simpler case (static controller) shown in Fig. \ref{fig:sf_scheme_classical}, with $D_c$ and $K$ being equivalent. When this is the case, we could still have a dynamic controller, but the dynamics (and hence stability) of the controller itself does not play any role on the stability of the control loop, allowing the controller to be unstable, until it stabilizes the plant.

Since in this article we deal with a dynamic system with process and measurement noises, deterministic notions of stability need to be extended accordingly. In particular, we will focus on the concept of MSS (mean square stability) \cite{willems1973mean}, where the system described on Eq. \eqref{eq:AandB} and \eqref{eq:CandD} is mean square stable if and only if, there exist $\mu_x \in \mathbb{R}^{n_x}$ and $M_x \in \mathbb{R}^{n_x\times n_x}$, $M_x \geq 0$, such that
\begin{subequations}\label{eq:mss}
\begin{align}
\lim_{t\rightarrow \infty}{\xi\{x(t)\}} &= \mu_x,\\
\lim_{t\rightarrow \infty}{\xi\{x(t)x(t)^{\intercal}\}} &= M_x.
\end{align}
\end{subequations}

This translates to the system being mean square stable if and only if its state $x$ has well defined and finite stationary mean and stationary second order moments matrix.

Theoretically there is no problem about having an unstable controller that stabilizes the control loop, but practically this represents a non-desirable choice, since in practical applications such as in robotics, an unstable controller could lead the physical system (robot respectively) to crash or get several damages.

Regarding the dimension of parameters of the state space model of the controller, we have $A_c \in \mathbb{R}^{n_{x_c}\times n_{x_c}}$, $B_c \in \mathbb{R}^{n_{x_c}\times nx}$, $C_c \in \mathbb{R}^{n_u\times n_{x_c}}$ and $D_c \in \mathbb{R}^{n_u\times n_x}$, where $n_{x_c}$ is the number of elements of the internal state of the controller $x_c(t)$, stressing that is not necessary to have $n_{x_c} = n_x$.

If we set the reference $r(t) = 0 \quad \forall t$, and $C_c = 0$, the problem reduces to regulation. The main objective of the regulation problem is to make the state $x(t)$ decrease to zero as $t \rightarrow \infty$, that is the reason for having $r(t) = 0$.

Consider a strictly causal plant described by Eq. \eqref{eq:AandB} and \eqref{eq:CandD} with $D=0$, and a controller given by Eq. \eqref{eq:controller}. Then, a MSS controller will stabilize the plant, in mean-square sense according to \eqref{eq:mss}, if and only if the eigenvalues of $\bar{A}$ lie on the left half-plane for the continuous-time case (and inside the unit circle for discrete-time case), where $\bar{A}$ is a block matrix given by
\begin{equation}\label{eq:Abar}
\bar{A} = \left[ \begin{array}{cc}A-BD_c & -BC_c\\ B_c & A_c\end{array}\right].
\end{equation}
If $\bar{A}$ is neither symmetric nor triangular, we can not obtain sufficient and necessary conditions on eigenvalues of matrices that compose $\bar{A}$. Nevertheless, eigenvalues of such block matrix would be given by:
\begin{equation}
eig(A-BD_c) \cup eig(A_c+B_c(A-BD_c)^{-1}BC_c),
\end{equation}
where making a slight abuse of notation, $eig(x)$ denotes the dominant eigenvalue of $x$.

Consider an augmented state vector $\bar{x}(t) = \left[\begin{array}{cc}x^{\intercal}(t) & x_c^{\intercal}(t)\end{array}\right]^{\intercal}$. Then, the augmented system is given by

\begin{align}\label{eq:augmented}
\bar{\dot{x}}(t) = \underbrace{\left[\begin{array}{cc}A-BD_c & -BC_c\\ B_c & A_c\end{array}\right]}_{\bar{A}}\bar{x}(t) &+ \underbrace{\left[\begin{array}{cc}B\\ 0_{nx\times nu}\end{array}\right]}_{\bar{B}}r(t)\\ \nonumber &+ \underbrace{\left[\begin{array}{cc}v(t)\\ 0_{nx_c\times 1}\end{array}\right]}_{\bar{v}(t)},
\end{align}

where $0_{n_{x_c}\times 1}$ is a zero-entries column vector with $n_{x_c}$ elements.

Then, discretizing the model with zero-order hold sampling and unitary sampling time, unrolling the system on Eq. \eqref{eq:augmented} in terms of the initial (augmented) state, yields
\begin{equation}
\bar{x}(t+1) = \bar{A}^{t+1}\bar{x}(0) + \sum_{i = 0}^{t}{\bar{A}^{i}\left(\bar{B}\;r(t-i) + \bar{v}(t-i)\right)},
\end{equation}
where given that $v(t)$ is zero-mean white noise and its expectation is given by
\begin{equation}
\mathcal{E}\{\bar{x}(t+1)\} = \mathcal{E}\{\bar{A}^{t+1}\}\bar{x}(0) + \sum_{i=0}^{t}{\mathcal{E}\{\bar{A}^i\bar{B}\;r(t-i)\}},
\end{equation}
that corresponds to a matrix power series, so for having finite $\mathcal{E}\{\bar{x}(t+1)\}$ in the limit when $t \rightarrow \infty$, $\bar{A}$ has to be such that
\begin{equation}
\Vert \bar{A} \Vert_2 < 1,
\end{equation}
that in terms of its dominant eigenvalue, condition is turned into
\begin{equation}
\vert eig(\bar{A})\vert < 1.
\end{equation}
It can be seen that making an analogous analysis for the second order moment of $\bar{x}(t)$ yields the same condition for $\bar{A}$, given that the only term affecting variance of $\bar{v}(t)$ is $v(t)$ and its variance is already assumed to be finite.

Recall that eigenvalues have a direct relationship with the speed i.e. the time-response achieves stationarity. Indeed, larger the eigenvalues (remain inside the unit circle), the transient of the time-response will disappear faster. If on the other hand, the eigenvalues are located outside the unit circle, the time-response will oscillate. Then, assume for a moment that $B_c = 0$ or $C_c = 0$, so it is clear to see that the controller can not be internally unstable and $D_c$ has to be designed such that $(A-BD_c)$ have eigenvalues inside the unit circle, otherwise stability of the loop and of the controller itself can not be guaranteed.

\section{Bayesian Optimization} \label{sec:bayesian}
BO \cite{DBLP:conf/bnaic/ThomasPLCWB19} is a derivative free optimization approach for global optimization of expensive black-box function $f$. It is a class of sequential-model based optimization algorithms that uses past evaluations of the function to find the next point to sample.  

To understand its necessity, consider a generic maximization problem,
\begin{equation}
        \mathbf{x}^*\; =\;\arg\max_{\mathbf{x}\in \mathcal{X}}\;f(\mathbf{x}).
    \end{equation}

where $\mathbf{x}^*$ is the global optimizer $\mathbf{x}$ is chosen from $\mathcal{X}$ where $\mathcal{X} \subset \mathbb{R}^m$ is a bounded domain. 

Variants of Bayesian optimization use different types of acquisition functions to determine the next point to be evaluated based on the current posterior distribution over functions. 

    The surrogate model used for this optimization is a Gaussian Processes (GP). A GP is characterized by its mean $\mu(\mathbf{x})$, and co-variance function $k (\mathbf{x},\mathbf{x}^{'})$. For $n$ data points, the function $f_{1;n}\;=\; f (\mathbf{x}_{1}), \;\ldots,\; f (\mathbf{x}_{n})$ can be characterized using a multivariate Gaussian distribution as
\begin{equation}
    f_{1:n}\;=\; \mathcal{N}\;(\mu\;(\mathbf{x}_{1:n}),\;\mathbf{K}),
\end{equation}
where $\mathbf{K}$ is a $n\times n$ kernel matrix given by
\begin{equation}
\mathbf{K}\;=\; \begin{pmatrix}
k(\mathbf{x}_1,\mathbf{x}_1) & \;\ldots & \; k(\mathbf{x}_1,\mathbf{x}_n) \\
\;\vdots & \;\ddots & \;\vdots\\
\;k(\mathbf{x}_n,\mathbf{x}_1) & \;\ldots & \;k(\mathbf{x}_n,\mathbf{x}_n)
\end{pmatrix},
\end{equation}
for some positive definite kernel such as a Gaussian or Matern kernel \cite{Rasmussen:2005:GPM:1162254}.

An acquisition function proposes that points should be selected next to determine the minimizer of the function, and they trade off between exploitation and exploration. We focus on the sequel on Expected Improvement \cite{BrochuarXiv2010} in this paper.

Expected Improvement evaluates $f(\mathbf{x})$ at the point where the expectation of the improvement in $f(\mathbf{x}^{+})$ under the current estimate of the GP is the highest
\begin{equation}
   \begin{aligned}[c]
EI(\mathbf{x})\; =\; &\EX\;[\max\{0\;,\; f(\mathbf{x}^{+}) \;-\;f(\mathbf{x})\}], \nonumber \\
 =\;  &(\mu(\mathbf{x})\;-\; f(\mathbf{\mathbf{x}}^{+})\;-\;\xi)\; \nonumber \\
 &\Phi \left( \frac{\mu(\mathbf{x})\;-\; f(\mathbf{x}^{+})\;-\;\xi}{\sigma(\mathbf{x})} \right)\; + \nonumber \\
 &\sigma(\mathbf{x})\; \phi \left( \frac{\mu(\mathbf{x})\;-\; f(\mathbf{x}^{+})\;-\;\xi}{\sigma(\mathbf{x})} \right) , \label{eq:ExpectedImprovementAcquisition}
\end{aligned} 
\end{equation}
where $f(\mathbf{x}^{+})$ is the best observed value of the function so far, $\mu(\mathbf{x})$ is the posterior mean of $\mathbf{x}$ under the GP, $\sigma(\mathbf{x})$ is the posterior standard deviation, and $\xi$ is a parameter used to drive exploration, that is usually very small. $\Phi$, $\phi$ are the cumulative distribution function and the probability density function of a standard normal variable respectively.

The algorithm for suspension design using Bayesian Optimization \cite{NoearXiv2018} is given in Algorithm \ref{alg1}.
\begin{algorithm}[h]{
  \SetAlgoLined
 \caption{Bayesian Optimization}
\label{alg1}
  \KwData{
  
  Initial Design $\mathbb{D}_{n_{init}}= \{(\mathbf{x}_i, y_i)\}_{i=1}^{n_{init}}$
  
  $n_{max}$ function evaluations} 
  \KwResult{
  
  Estimated max : 
  \begin{equation*}
    f_{max} = \max{(|f(\mathbf{x}_1)|,\;\ldots,\; |f(\mathbf{x}_{n_{max}})|)}
  \end{equation*}
  
  Estimated maximum point :
  \begin{equation*}
  \mathbf{x}_{max} =  \argmax{(|f(\mathbf{x}_1)|,\;\ldots,\; |f(\mathbf{x}_{n_{max}})|)}
  \end{equation*}
  }
  \For {$n = n_{init}$ to $n_{max}$}{
  
  Update GP: 
  \begin{equation*}
      f(\mathbf{x}) | \mathbb{D}_n \sim GP(\hat{f}(\mathbf{x}), \mathbf{K}(\mathbf{x},\mathbf{x^{'}}))
  \end{equation*}
  
   Optimize acquisition function:
   \begin{equation*}
    \mathbf{x}_{next} = \argmax_{\mathbf{x} \in \mathcal{X}} \alpha_n(\mathbf{x})
   \end{equation*}
   
  Find $f$ at $\mathbf{x_{next}}$ to obtain $y_{next}$
  
  Add data to previous design:
  \begin{equation*}
    \mathbb{D}_{n+1} = \mathbb{D}_n \cup \{\mathbf{x}_{next},y_{next}\}
  \end{equation*}
  }
}
\end{algorithm}

\section{Experimental Results} \label{sec:results}
\subsection{Toy Example}
First, consider a discrete-time SISO plant (one-input one-output, $n_u = 1, n_y = 1$) be given by
\begin{subequations}
\begin{align}
x[k+1] &= Ax[k] + Bu[k] + v[k],\\
y[k] &= Cx[k] + w[k],
\end{align}
\end{subequations}
where as before, $v[k]$ and $w[k]$ are the process and measurement noise respectively, zero-mean and unitary variance white noises.

Let $A$, $B$, $C$ and $D$ be given by
\begin{subequations}
\begin{align}
A &= \left[\begin{array}{cc}0.5 & 0\\ 0.7 & 1.2\end{array}\right],\\
B &= \left[\begin{array}{cc}0\\ 0.1\end{array}\right],\\
C &= \left[\begin{array}{cc}1 & 1\end{array}\right],\\
D &= 0.
\end{align}
\end{subequations}

It can be observed that the plant is internally unstable, since one of its eigenvalues lie outside the unit circle. Then, parameters $A_c$, $B_c$, $C_c$, and $D_c$ of the controller have to be found for generating a (scalar) control signal $u[k]$,
\begin{subequations}\label{eq:cont}
\begin{align}
x_c[k+1] &= A_cx_c[k] + B_cx[k],\\
u[k] &= r[k] -\left(C_cx_c[k] + D_cx[k]\right),
\end{align}
\end{subequations}
whose matrices were set to
\begin{align*}
A_c = 0.4 &\qquad B_c = \left[\begin{array}{cc}1 & -1.52\end{array}\right],\\
C_c = -0.5 &\qquad D_c = \left[\begin{array}{cc}0.3 & 2.1\end{array}\right].
\end{align*}
These values were chosen such that the augmented system matrix $\bar{A}$ has prescribed eigenvalues. Indeed, $eig(\bar{A}) \in \{0.5, 0.59, 0.8\}$.

Note from Eq. \eqref{eq:cont}, that we have assumed for this first example that the feedback is made over the true state. The reference has been set to
\begin{equation}
r[k] = \begin{cases}
0 \qquad \text{ $k < 60$}\\
10 \qquad \text{$k \geq 60$}
\end{cases}
\end{equation}

We compare this control scheme with the classical state-feedback architecture shown in Fig. \ref{fig:sf_scheme_classical}, with output $y[k]$, such that the control signal is given by
\begin{equation}
u[k] = r[k]-Kx[k],
\end{equation}
and $K$ is set to
\begin{equation*}
K = \left[\begin{array}{cc}0.3 & 4\end{array}\right],
\end{equation*}
such that $eig(A-BK) \in \{0.5, 0.8\}$.

As expected, from analyzing eigenvalues of $\bar{A}$ and $A-BK$, we see that the closed loop system achieves stationarity with the same velocity (given by its dominant eigenvalue, that represents the velocity of the slowest disappearing natural mode), but with different values for $D_c$ and $K$. This is possible because in Fig. \ref{fig:sf_scheme}, we added degrees of freedom on the controller with respect to the scheme shown in Fig. \ref{fig:sf_scheme_classical}, at the cost of introducing an additional natural mode on the closed loop response.

\begin{figure}[ht]
\centering
\includegraphics[width=.5\textwidth]{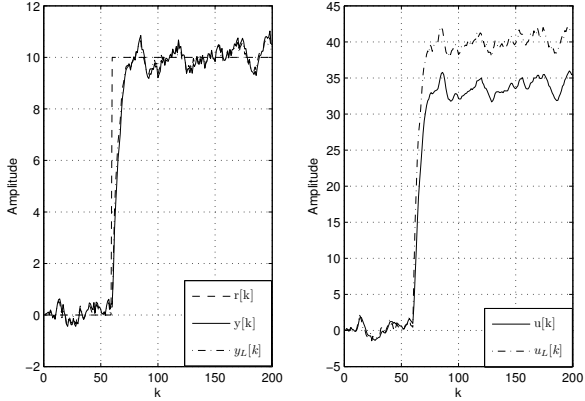}
\caption{Full state-feedback, step reference tracking}
\label{fig:stepsimulation}
\end{figure}

As shown on simulation results depicted on Fig. \ref{fig:stepsimulation}, the introduced degrees of freedom on the controller allow us to set matrix values such that the closed loop response achieves stationarity as fast as desired, and confining the control signal to be smaller than in the classical architecture.

\begin{figure}[ht]
\centering
\includegraphics[width=.5\textwidth]{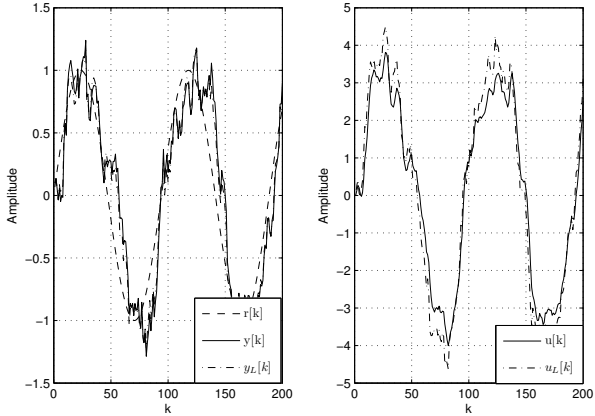}
\caption{Full state-feedback, sinusoidal reference tracking}
\label{fig:sinesimulation}
\end{figure}

Fig. \ref{fig:sinesimulation} shows the same comparison for a sinusoidal reference signal, where the results remain the same, namely, speed of convergence is the same although the dynamic controller achieves lower peaks on the control signal, despite the difference on the values of $D_c$ and $K$, due to the additional degrees of freedom. Nevertheless, the reader should note that reference tracking in this case will be achievable just if the closed loop dynamics are fast enough to keep track of changes on the reference.

\subsection{Rotary Inverted Pendulum}
When considering parameters on Table \ref{tab:parameters} for model described on Eq. \eqref{eq:AandB} and \eqref{eq:CandD}, following matrices are obtained:

\begin{subequations}\label{eq:ripmodel}
\begin{align}
    A &= \left[\begin{array}{ccccc} 0 & 1 0 & 0& 0\\ 0 & -0.31 & 2.99 & -0.02 & 81.47\\ 0 & 0 & 0 & 1 & 0\\ 0 & -0.02 & 61.49 & -0.5& 63.88\\ 0 & -261.94 & 0 & 0 & -800\end{array}\right],\\
    B &= \left[\begin{array}{ccccc}0&0&0&0&100\end{array}\right]^{\intercal},\\
    C &= \left[\begin{array}{ccccc}1&0&0&0&0\\0&0&1&0&0\\0&0&0&0&1\end{array}\right].
\end{align}
\end{subequations}

Performance of proposed controller on Eq. \eqref{eq:controller} for the Rotary Inverted Pendulum will be compared with the classical state-feedback controller on Eq. \eqref{eq:lqr}. To this end, matrices $Q$ and $R$ on Eq. \eqref{eq:costlqr} have to be designed by means of giving more weight to states that are more relevant than others for the particular experiment. In this case, priority is given to pendulum position, $\theta_1$, while other important variables correspond to arm and pendulum velocities, so experiment proposal is given by

\begin{subequations}
\begin{align}
    Q &= \left[\begin{array}{ccccc}1&0&0&0&0\\0&10&0&0&0\\0&0&100&0&0\\0&0&0&10&0\\0&0&0&0&1\end{array}\right],\\
    R &= 10,
\end{align}
\end{subequations}
where penalizing actuation factor, $R$, is chosen such that DC motor does not perform any sudden behavior changes destabilizing pendulum position.

Then, solving Eq. \eqref{eq:are} with parameters described on Eq. \eqref{eq:ripmodel} yields:

\begin{equation}
    K = \left[\begin{array}{ccccc}-0.31 & -5.26 & 70.74 & 8.92 & 0.17\end{array}\right].
\end{equation}

On the other hand, when considering controllers to be of the form described on Eq. \eqref{eq:controller}, parameters are obtained by means of applying Bayesian Optimization as detailed in Algorithm \ref{alg1} with 150 maximum iterations and minimizing:
\begin{equation}
    \vert \mathrm{Re}\{eig(\bar{A})\}\vert_{\infty},
\end{equation}
where $\bar{A}$ is the same on Eq. \eqref{eq:Abar}, $\mathrm{Re}\{eig(M)\}$ stands for the real part of eigenvalues from matrix $M$. Then, minimization of infinite norm is used for constraining maximum eigenvalue and therefore dominant velocity, that leads to:

\begin{subequations}
\begin{align}
    A_c &= -100,\\
    B_c &= \left[\begin{array}{ccccc}0&0&0&0&0\end{array}\right],\\
    C_c &= -0.5,\\
    D_c &= \left[\begin{array}{ccccc}-20.96 & -39.76 & 72.74 & 92.61 & -0.58\end{array}\right].
\end{align}
\end{subequations}

\begin{figure*}[!htbp]
    \centering
    \includegraphics[width=1.1\textwidth]{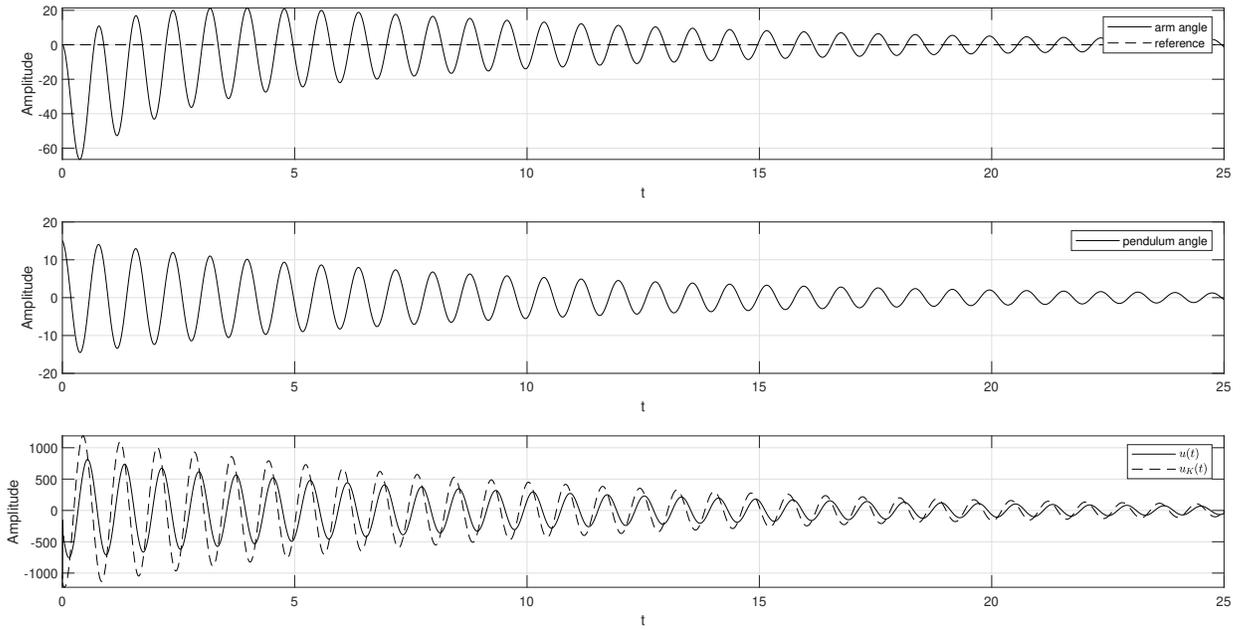}
    \caption{Furuta pendulum control}
    \label{fig:comparison}
\end{figure*}

\begin{figure*}[!htbp]
    \centering
    \includegraphics[width=1.1\textwidth]{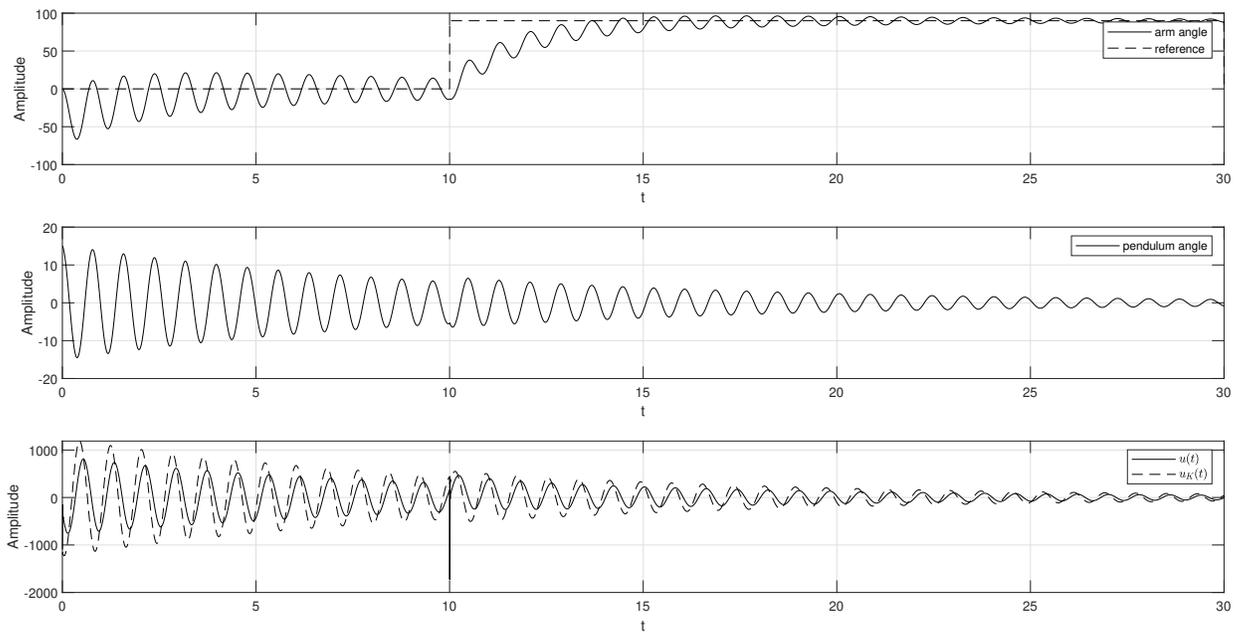}
    \caption{Furuta pendulum control with reference tracking for arm angle}
    \label{fig:trackingcomparison}
\end{figure*}

Fig. \ref{fig:comparison} shows on the upper and middle subplots the arm and pendulum position respectively, observing that both types of controllers are capable of stabilizing the pendulum position at the same time that arm position is being controlled. Lower subplot shows the same behavior obtained on Fig. \ref{fig:sinesimulation}, where controller on Eq. \eqref{eq:controller} achieves the same performance of Eq. \eqref{eq:lqr} but with smaller actuation magnitude. Fig. \ref{fig:trackingcomparison} also shows this behavior when reference for the arm position is changed while the pendulum position is being controlled at the same time. The implementation of the complete paper is available at \href{https://github.com/miguel-a-solis/Solis-Thomas2021}{https://github.com/miguel-a-solis/Solis-Thomas2021}.

\section{Conclusion} \label{sec:conclusion}
This work has documented a Bayesian Optimization approach for tuning a generalized state-feedback controller, and its performance comparison against the classical controller obtained by solving the Linear Quadratic Regulator (LQR) problem. Simulation results showed that the proposed controller scheme performs the same as the classical one in terms of the tracking error, but this is achieved at a lower effort in terms of the control signal.

Also, a well-known underactuated system, the rotary inverted pendulum (also known as Furuta pendulum) was studied and a linear model was obtained. Simulation results were also obtained on this plant, and again performance was compared against the classical state-feedback controller, also showing that tracking performance was equivalent but at lower control effort. Experimental results were obtained under a simulated environment, while implementation on the physical prototype remains as future work.

\bibliographystyle{IEEEtran}
\bibliography{references}

\end{document}